\documentclass[twocolumn]{webofc}
\usepackage[varg]{txfonts}   % Web of Conferences font
\usepackage{hyperref}
\usepackage{url}
\hypersetup{colorlinks=true,citecolor=blue,urlcolor=blue,linkcolor=blue}
\usepackage{graphicx}  % include figure files
\usepackage{float}
\usepackage{epstopdf}
\usepackage{amsmath}
\usepackage{amssymb}  % for the use of \blacksquare
\usepackage{wasysym}  % for the use of \CIRCLE
\usepackage{xcolor,xspace}
\usepackage{braket}
\usepackage[percent]{overpic}  % insert LaTex text over 
\newcommand{\ba}{\begin{eqnarray}}
\newcommand{\ea}{\end{eqnarray}}
\newcommand{\bsub}{\begin{subequations}}
\newcommand{\esub}{\end{subequations}}

\def\ket#1{|#1\rangle}

\def\beq{\beta_{\rm eq}}
\def\g0{\gamma_0}

\begin{document}
%
%\title{Insert your title here}
\title{Quantum and classical analyses of intertwined
  phase transitions\\
  in odd-mass Nb isotopes}

\author{\firstname{A.} \lastname{Leviatan}\fnsep\thanks{\email{ami@phys.huji.ac.il}}
}

\institute{Racah Institute of Physics, The Hebrew University, 
Jerusalem 91904, Israel
          }

%\author{\firstname{Solange} \lastname{Gu\'ehot}\inst{1,3}\fnsep\thanks{\email{Mail address for first
%    author}} \and
%        \firstname{Isabelle} \lastname{Houlbert}\inst{2}\fnsep\thanks{\email{Mail address for second
%             author if necessary}} \and
%        \firstname{AgnÃ¨s} \lastname{Henri}\inst{3}\fnsep\thanks{\email{Mail address for last
%             author if necessary}}
%        % etc.
%}

%\institute{Insert the first address here 
%\and
%           the second here 
%\and
%           Last address
%          }

\abstract{
  Quantum phase transitions (QPTs) in odd-mass Nb
  isotopes are investigated in the framework
  of the interacting boson-fermion model
  with configuration mixing.
  A quantum analysis reveals a Type~I QPT
  (gradual shape-evolution within the intruder
  configuration)  superimposed on a Type~II QPT (abrupt
  crossing of normal and intruder states), thus
  demonstrating the occurrence of intertwined QPTs.
  A classical analysis highlights the implications
  for the single particle motion in the deformed field
  generated by the even-even Zr cores.
}
\maketitle
\section{Introduction}
\label{intro}
Quantum phase transitions (QPTs) are
qualititative changes in the
properties of a physical system induced by variation of
parameters in the Hamiltonian. Such structural changes
are manifested empirically in nuclei as transistions
between shapes and can be categorized into two types.
The first, denoted as 
\mbox{Type~I}~\cite{Diep80}, is a shape-phase
transition within a single configuration, as encountered
in the neutron number~90 region~\cite{Cejnar2010}.
The second, denoted as \mbox{Type~II}~\cite{Frank2006},
is a phase transition involving a crossing of different 
configurations, as encountered
in nuclei near (sub-) shell closure~\cite{Heyde11}.
In most cases, the strong mixing between the
configurations obscures the individual QPTs.
However, if the mixing is small,
the Type~II QPT can be accompanied by a distinguished
Type~I QPT within each configuration separately.
Such a scenario, referred to as intertwined QPTs,
occurs in the even-even Zr~(Z=40)
isotopes~\cite{Gavrielov2019,Gavrielov2020,Gavrielov2022}.
In the present contribution, we discuss both quantum
and classical analyses of a similar scenario
in the adjacent odd-even Nb~(Z=41)
isotopes~\cite{gavleviac22,Gav23,LevGav25}.

\vspace{-0.02cm}
\section{IBFM for multiple configurations}
\label{sec-IBFMCM}
Odd-A nuclei are treated in the interacting boson-fermion
model (IBFM)~\cite{IBFMBook}, as a system of monopole ($s$) 
and quadruple ($d$) bosons, representing valence nucleon 
pairs, and a single (unpaired) nucleon.
The (single-configuration) IBFM Hamiltonian has the form
\ba
\hat H(\xi) = \hat H_{\rm b} + \hat H_{\rm f}
+ \hat V_{\rm bf} ~.
\label{Hibfm}
\ea
We focus the discussion to
an odd fermion in 
a single-$j$ orbit for which,
\bsub
\ba
&&
\hspace{-1cm}
\hat H_{\rm b}(\epsilon_d,\kappa,\chi) =
  \epsilon_d\,
  \hat n_d + \kappa\,
  \hat Q_\chi \cdot \hat Q_\chi
\label{Hb}\\
&&
\hspace{-1cm}
\hat{H}_{\rm f}(\epsilon_j) = \epsilon_{j}\,\hat{n}_j
\label{Hf}\\
&&
\hspace{-1cm}
\hat{V}_{\rm bf}(\chi,A,\Gamma,\Lambda) =
A\,\hat{n}_d\,\hat{n}_j + \Gamma\,\hat{Q}_{\chi}\cdot
( a_{j}^{\dag }\, \tilde{a}_{j} )^{(2)}
\nonumber\\
&&
+ \Lambda
\sqrt{2j+1}:[ ( d^{\dag }\, \tilde{a}_{j})^{(j)}\times
  ( \tilde{d}\, a_{j}^{\dag })^{(j)}]^{(0)}:
\label{Vbf}
\ea
\label{Hibfm1}
\esub
The boson part~(\ref{Hb}) contains pairing and
quadrupole terms,
where $\hat{n}_d\!=\!\sum_{m}d^{\dag}_md_m$,
$\hat Q_\chi \!=\!
d^\dag s+s^\dag \tilde d +\
\chi\, (d^\dag \tilde d)^{(2)}$ and
$\tilde{d}_m\!=\!(-1)^{m}d_{-m}$.
The fermion part~(\ref{Hf}) involves the
number operator, $\hat{n}_j$,
and the boson-fermion part~(\ref{Vbf})
is composed of monopole,
quadrupole and exchange  terms, where
$:\!-\!:$
denotes normal ordering and
$\tilde{a}_{j,m} \!=\! (-1)^{j-m}a_{j,-m}$.
A change in the parameters~$\xi=
(\epsilon_d,\kappa,\chi,\epsilon_j,A,\Gamma,\Lambda$)
of $\hat{H}(\xi)$ induces Type~I
QPTs~\cite{Petrellis2011a,Petrellis2011b,Boyukata2021},
relevant to odd-mass nuclei.

The interacting boson-fermion model with configuration
mixing (IBFM-CM)~\cite{gavleviac22,Gav23} is an
extension of the IBFM to include
configurations with $N,N\!+\!2,\dots$ bosons,
representing shell model spaces built on
0p-0h, 2p-2h,$\ldots$ particle-hole
excitations across closed shells. For two
configurations ($A,B$), the IBFM-CM Hamiltonian
can be cast in matrix form,
\ba
\hat{H}(\xi_A,\xi_B,\omega) =
\begin{bmatrix}
  \hat{H}_A(\xi_A) & \hat{W}(\omega)\\
  \hat{W}(\omega) & \hat{H}_B(\xi_B)
\end{bmatrix} ~.
\label{Hibfm-cm}
\ea
Here $\hat{H}_A(\xi_A)$ and $\hat{H}_B(\xi_B)$
represent, respectively, the normal
$A$~configuration ($N$ boson space)
and the intruder $B$~configuration ($N+2$ boson space),
both coupled to a single $j$-fermion and $\hat{W}(\omega)$
a mixing term. Specifically,
\bsub
\ba
&&
\hspace{-1.5cm}
\hat{H}_A(\xi_A) =
\hat H_{\rm b}(\epsilon^{A}_d,\kappa_{A},\chi)
+ \hat{H}_{\rm f} + \hat{V}_{\rm bf} ~,\\
&&
\hspace{-1.5cm}
\hat{H}_B(\xi_B) = 
\hat H_{\rm b}(\epsilon^{B}_d,\kappa_{B},\chi) + 
\kappa^{\prime}_{B}\, \hat L \cdot \hat L + \Delta_B
+ \hat{H}_{\rm f} + \hat{V}_{\rm bf} ~,\\
&&
\hspace{-1.5cm}
\hat W(\omega) =
    \omega\, [\,(d^\dag d^\dag)^{(0)} + (s^\dag)^2\,
      + {\rm H.c.} ] ~.
\ea
\label{Hibfm-cm2}
\esub
where $\hat L \cdot \hat L$ is a rotational term,
$\Delta_B$ an energy off-set and H.c. stands for
Hermitian conjugate.
For simplicity, $\hat{H}_{\rm f}(\epsilon_j)$ and
$\hat{V}_{\rm  bf}(A,\Gamma,\lambda)$
are taken to be the same in both configurations.
A change in the parameters $(\xi_A,\xi_B,\omega)$
can induce a Type~I QPT within each configuration and
a Type~II QPT of configuration crossing. 
For $\hat{H}_{\rm f} = \hat{V}_{\rm bf} =0$,
the IBFM-CM Hamiltonian of Eq.~(\ref{Hibfm-cm}) reduces
to that of the interacting boson model with configuration
mixing (IBM-CM)~\cite{Duval1981, Duval1982},
relevant to the study of configuration-mixed
QPTs and shape-coexistence in even-even
nuclei.
\begin{figure}[t]
\centering
\includegraphics[width=\linewidth]{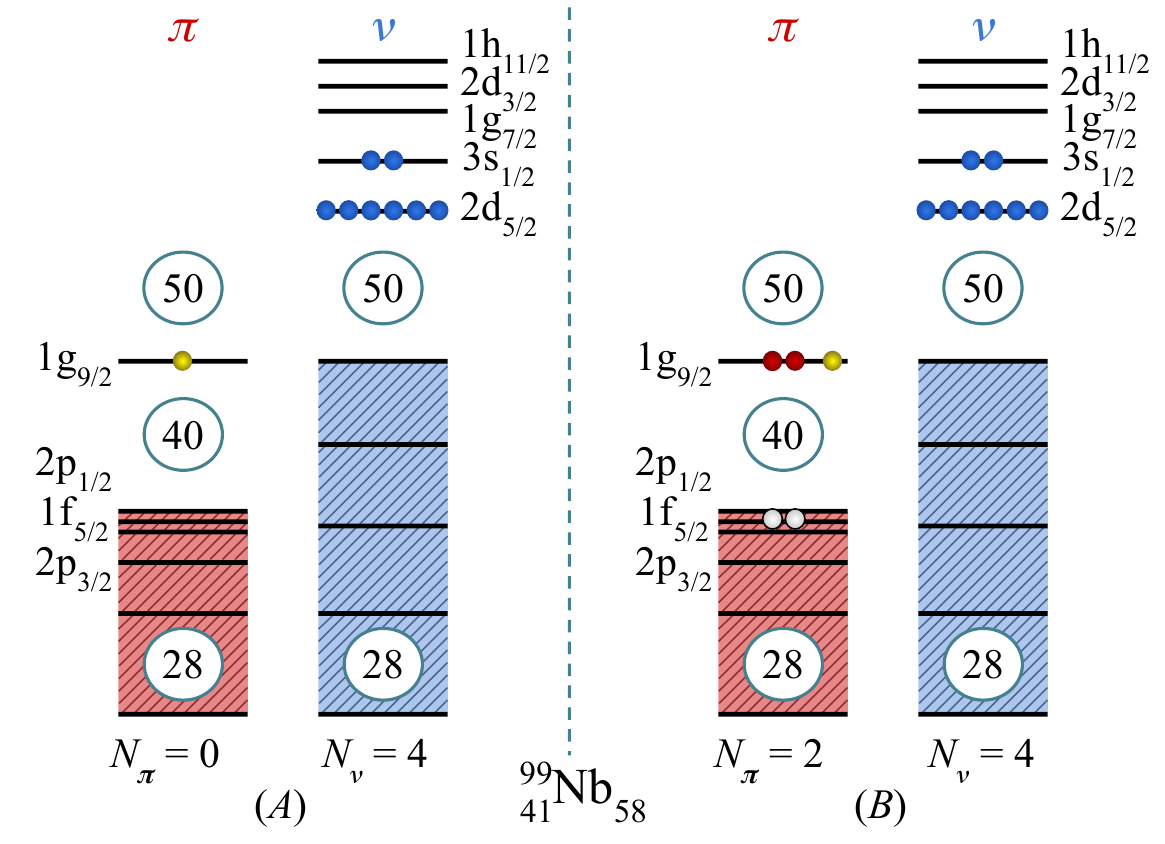}
\caption{Schematic representation of the two coexisting
  shell-model configurations ($A$ and $B$) for
  $^{99}_{41}$Nb$_{58}$. The corresponding numbers of
  proton bosons ($N_{\pi}$) and neutron bosons ($N_{\nu}$),
  are listed for each configuration and $N=N_{\pi}+N_{\nu}$.
\label{fig1-shells-99Nb}
}
\end{figure}

The eigenstates $\ket{\Psi;j,J}$
of $\hat{H}$, Eq.~(\ref{Hibfm-cm}),
are linear combinations of wave functions $\Psi_{\rm A}$ 
and $\Psi_{\rm B}$, involving bosonic basis states in the 
two spaces $\ket{[N],\alpha,L}$ and
$\ket{[N\!+\!2],\alpha,L}$, coupled to a $j$-fermion.
The boson ($L$) and fermion ($j$) angular momenta are 
coupled to $J$,
\ba
&&
\hspace{-1cm}
\ket{\Psi;j,J} =
a\,\ket{\Psi_A,[N],j;J} + b\,\ket{\Psi_B,[N\!+\!2],j;J}
\nonumber\\[2mm]
&&
\hspace{0.2cm}
= \sum_{\alpha,L}C^{(N,J)}_{\alpha,L,j}
\ket{N,\alpha,(L\otimes j)J}
\nonumber\\
&&
\hspace{0.5cm}
+ \sum_{\alpha,L}C^{(N+2,J)}_{\alpha,L,j}
\ket{N+2,\alpha,(L\otimes j)J} ~.
\ea
The probability of normal-intruder mixing is given by,
\begin{equation}
\label{Prob-ab-cm}
  a^2 =\sum_{\alpha,L}|C^{(N,J)}_{\alpha,L,j}|^2 \;\;\; ,
  \;\;\;
  b^2 = \sum_{\alpha,L}|C^{(N+2,J)}_{\alpha,L,j}|^2
= 1-a^2 ~.
\end{equation}

\section{Quantum analysis of the Nb chain}
\label{sec-Nb}
Positive-parity states in $^{A}_{41}$Nb isotopes
with mass number \mbox{$A\!=\!\text{93--105}$}
are described in the shell model by
coupling a proton in a $\pi(1g_{9/2})$ orbit to
the respective $_{40}$Zr cores with
neutron number 52--64.
In the latter, the normal 
A~configuration corresponds to having no active protons 
above the $Z\!=\!40$ sub-shell gap, and the intruder 
B~configuration corresponds to two-proton excitation from 
below to above this gap, creating 2p-2h states.
The IBFM-CM model space employed,
consists of a single $j=9/2$ fermion and
$[N]\oplus[N+2]$ boson spaces with
$N\!=\!1,2,\ldots,7$ for $^{93-105}$Nb.
The two configurations relevant for $^{99}$Nb
are shown schematically in Fig.~\ref{fig1-shells-99Nb}.
The parameters of the IBFM-CM
Hamiltonian~(\ref{Hibfm-cm2}) and transition
operators are determined from a fit in the manner
described in~\cite{gavleviac22,Gav23,LevGav25}.
\begin{figure}[t]
\centering
\begin{overpic}[width=\linewidth]{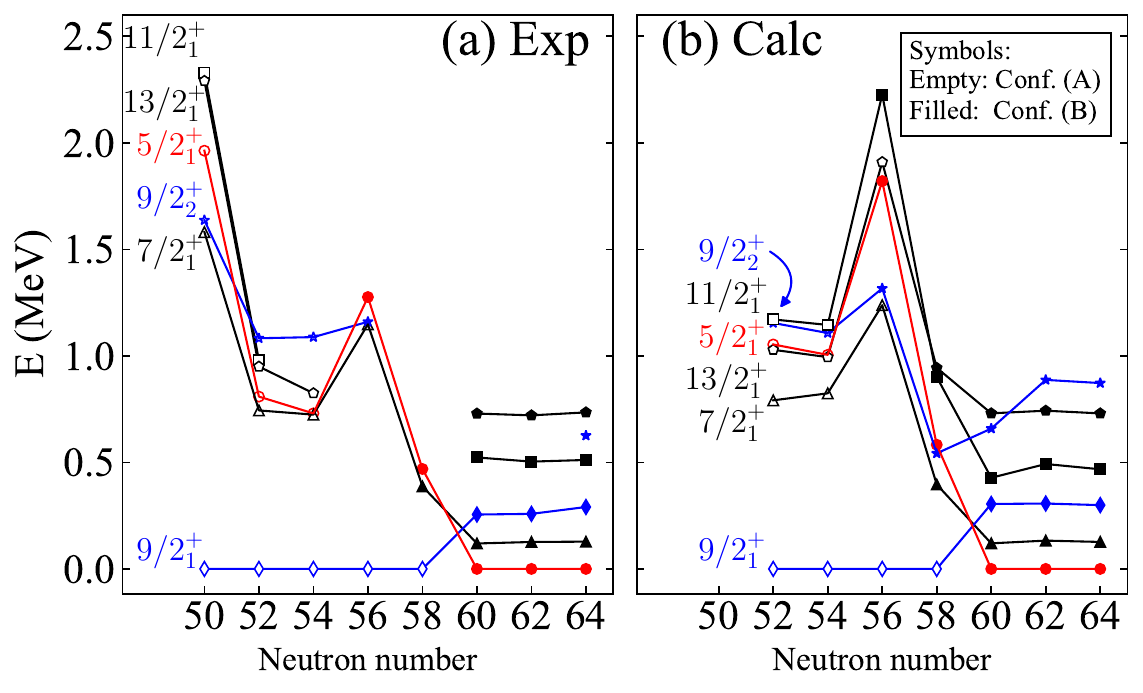}
\end{overpic}
\caption{Comparison between (a)~experimental and
(b)~calculated
  lowest-energy positive-parity levels in Nb isotopes.
  Empty (filled) 
symbols indicate a state dominated by the normal A 
(intruder B) configuration.
In particular, the $9/2^+_1$ state is in the A (B)
configuration for  
neutron number 52--58 (60--64)
and the $5/2^+_1$ state is  
in the A (B) configuration for 52--54 (56--64).
\label{fig:energies-p}}
\end{figure}
\begin{figure}[t]
\centering
\includegraphics[width=1\linewidth]{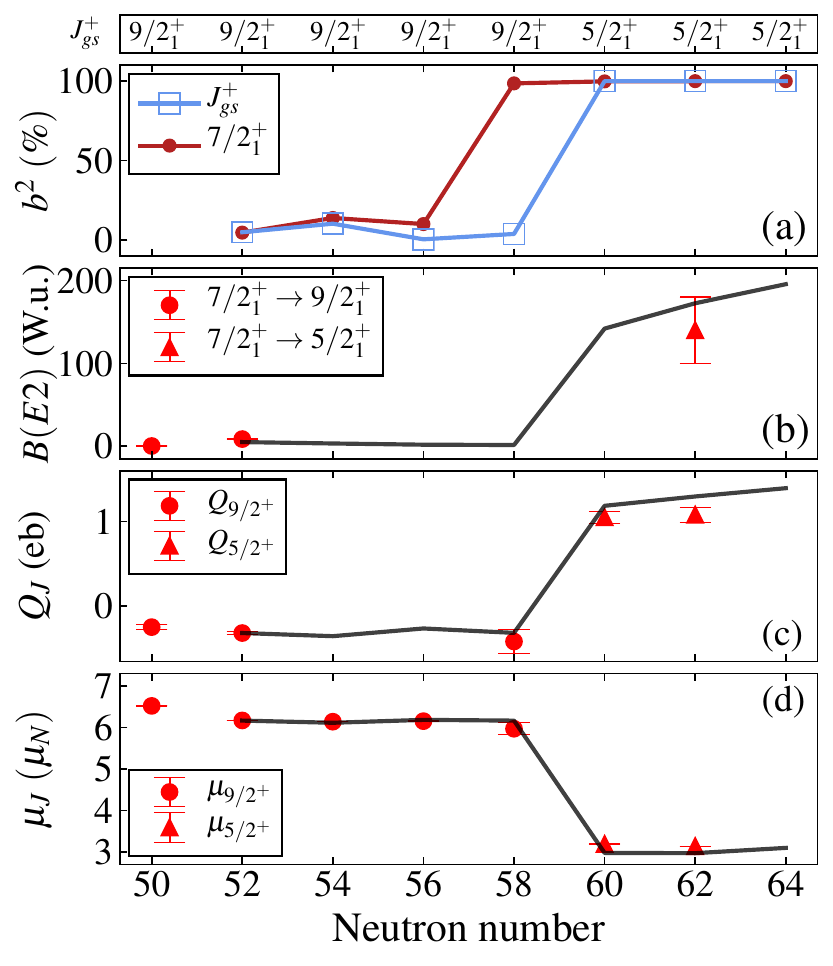}
\caption{\label{fig-b2-be2-q-mu}
  Evolution of spectral properties along the Nb~chain.
  Symbols (solid lines) denote experimental data 
  (calculated results). (a)~Percentage of the intruder (B)
  component [the $b^2$ probability in
    Eq.~(\ref{Prob-ab-cm})], in the ground state ($J^+_{gs}$)
and first-excited state ($7/2^+_1$).
Values of $J^+_{gs}$ are indicated at 
the top. (b)~$B(E2; 7/2^+_1\to J^{+}_{gs})$ in W.u. 
(c)~Quadrupole moments of $J^+_{gs}$ in $eb$. (d)~Magnetic 
moments of $J^+_{gs}$ in $\mu_N$.
For the data,
see Fig.~2 of~\cite{gavleviac22}.
\label{fig:order-params}}
\end{figure}

Figure~\ref{fig:energies-p}
shows the experimental
and calculated levels of selected states
in the Nb isotopes
along with assignments to configurations based on 
Eq.~(\ref{Prob-ab-cm}). Empty (filled) symbols 
indicate a dominantly normal (intruder) state with small 
(large) $b^2$ probability. In the region between neutron 
number 50 and 56, there appear to be two sets of levels 
with weakly deformed structure, associated with 
configurations A and B. All levels decrease in energy 
for 52--54, away from closed shell, and rise again 
at 56 due to the neutron $\nu(2d_{5/2})$ subshell closure.
From 58, there is a pronounced drop in energy for
the states of the B~configuration.
At 60, the two configuration cross, 
indicating a Type~II QPT, and the ground state changes
from $9/2^+_1$ to $5/2^+_1$, becoming the bandhead of a 
$K=5/2^+$ rotational band composed of  $5/2^+_1, 7/2^+_1, 
9/2^+_1, 11/2^+_1, 13/2^+_1$ states. The intruder 
B~configuration remains strongly deformed and the band 
structure persists beyond 60.
The above trend is similar to that encountered in the
even-even Zr isotopes with the same 
neutron numbers (see Fig.~14 of~\cite{Gavrielov2022}).

A possible change in the angular momentum of the ground 
state ($J^{+}_{gs}$) is a characteristic signature of 
Type~II QPTs in odd-mass, unlike in even-even nuclei
where the 
ground state remains $0^+$ after the crossing. It is an 
important measure for the quality of the calculations, 
since a mean-field approach, without configuration mixing, 
fails to reproduce the switch from $9/2^+_1$
to $5/2^+_1$ in $J^{+}_{gs}$ for the Nb
isotopes~\cite{Guzman2011}. 
Fig.~\ref{fig-b2-be2-q-mu}(a)
shows the percentage of the wave function within 
the B~configuration for $J^+_{gs}$ and $7/2^+_1$, as a 
function of neutron number across the Nb chain. The rapid 
change in structure of $J^+_{gs}$ from the normal 
A~configuration in $^{93-99}$Nb (small $b^2$ probability)
to the intruder B~configuration in $^{101-105}$Nb (large 
$b^2$) is clearly evident, signaling a Type~II QPT. The 
configuration change appears sooner in the $7/2^+_1$ state, 
which changes to the B~configuration already in $^{99}$Nb.  
Outside a narrow region near neutron number 60, where the 
crossing occurs, the two configurations are weakly mixed 
and the states retain a high level of purity.

\begin{figure}[t!]
\centering
\begin{overpic}[width=\linewidth]{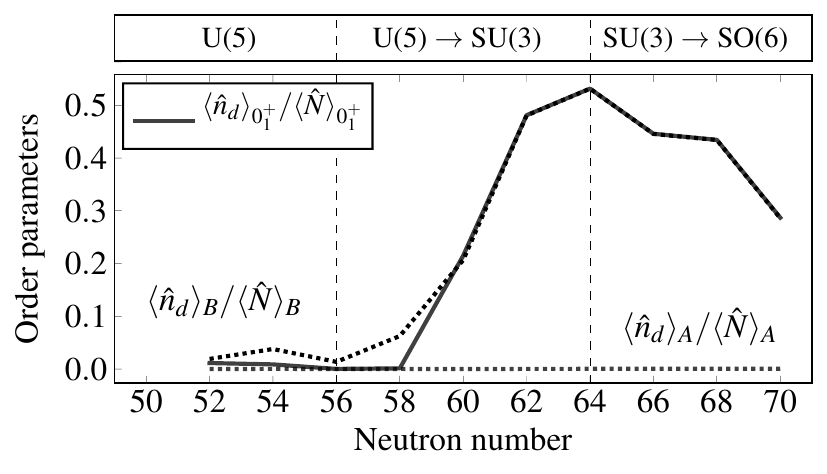}
\put(90,42) {\large (a)}
\end{overpic}\\
\begin{overpic}[width=\linewidth]{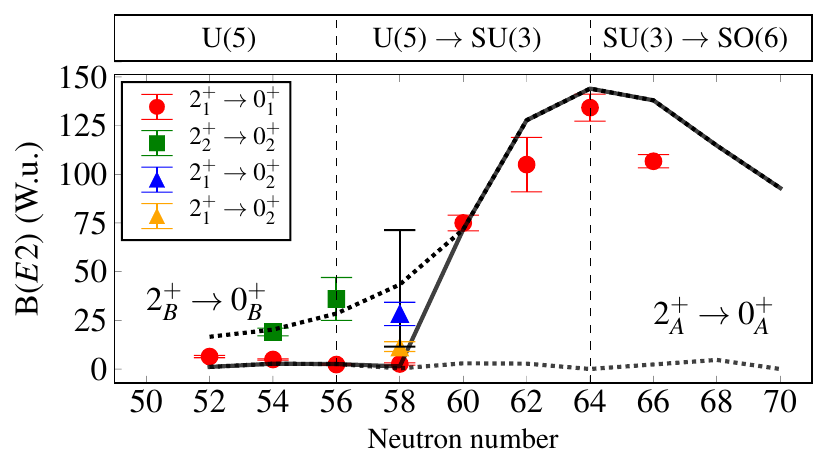}
\put(90,42) {\large (b)}
\end{overpic}
\caption{\label{fig-nd-be2-Zr}
  Evolution of spectral properties along the Zr chain.
  (a)~Order parameters: $\braket{\hat n_d}_{0^+_1}$
(solid line) and
$\braket{\hat n_d}_{A},\,\braket{\hat n_d}_{B}$
(dotted lines), normalized by the respective
boson numbers.
  (b)~B(E2) values for $2^+\!\to\!0^+$ transitions
  in Weisskopf units (W.u.). Dotted
  lines denote calculated $E2$ rates for transitions
  within a  configuration. Solid line denotes
  calculated $2^{+}_1\to 0^{+}_1$ rates.
  For the data, see Fig.~17 of~\cite{Gavrielov2022}.
}
\end{figure}

Further insight into the nature of the QPTs is gained by
considering the behaviour of the order parameters and
related observables. The $B(E2; 7/2^+_1\!\to\! J^{+}_{gs})$
and quadrupole moment of $J^{+}_{gs}$
in Nb isotopes are shown
in Fig.~\ref{fig-b2-be2-q-mu}(b)
and Fig.~\ref{fig-b2-be2-q-mu}(c), respectively.
These observables are related to the deformation, the order 
parameter of the QPT. Although the data is incomplete, one 
can still observe small (large) values of these observables 
below (above) neutron number 60, indicating an increase in 
deformation. The calculation reproduces well this trend and 
attributes it to a Type~II QPT involving a jump between 
neutron number 58 and 60, from a weakly-deformed 
A~configuration, to a strongly-deformed B~configuration. 
Such a Type~II scenario is supported also by the trend in 
the magnetic moments ($\mu_J$) of the ground state, shown 
in~\ref{fig-b2-be2-q-mu}(d), where both the data and the 
calculations show a constant value of $\mu_J$ for neutron 
numbers 52--58, and a drop to a lower value at 60, which 
persists for 60--64. The behavior shown in
Figs.~\ref{fig-b2-be2-q-mu}(b)-\ref{fig-b2-be2-q-mu}(c)
is correlated with 
a similar jump seen for the 
B(E2)'s of $2^+\!\to\!0^+$ transitions in the even-even
Zr cores, Fig.~\ref{fig-nd-be2-Zr}(b), and with the
calculated order
parameters, Fig.~\ref{fig-nd-be2-Zr}(a). The latter are
the expectation value of
$\hat{n}_d$ in the $0^{+}_1$ ground state wave function,
$\braket{\hat{n}_d}_{0^{+}_1}$,
and in its normal ($A$) and intruder ($B$) components,
$\braket{\hat{n}_d}_A$ and $\braket{\hat{n}_d}_B$.
Their evolution along the Zr chain reveals that
configuration~$A$ remains spherical, while
configuration~$B$
undergoes a Type~I QPT involving a gradual
spherical-to-deformed [U(5)-SU(3)-SO(6)] shape-phase
transition~\cite{Gavrielov2019,Gavrielov2020,Gavrielov2022}.
\begin{figure*}
\centering
\includegraphics[width=1\linewidth]{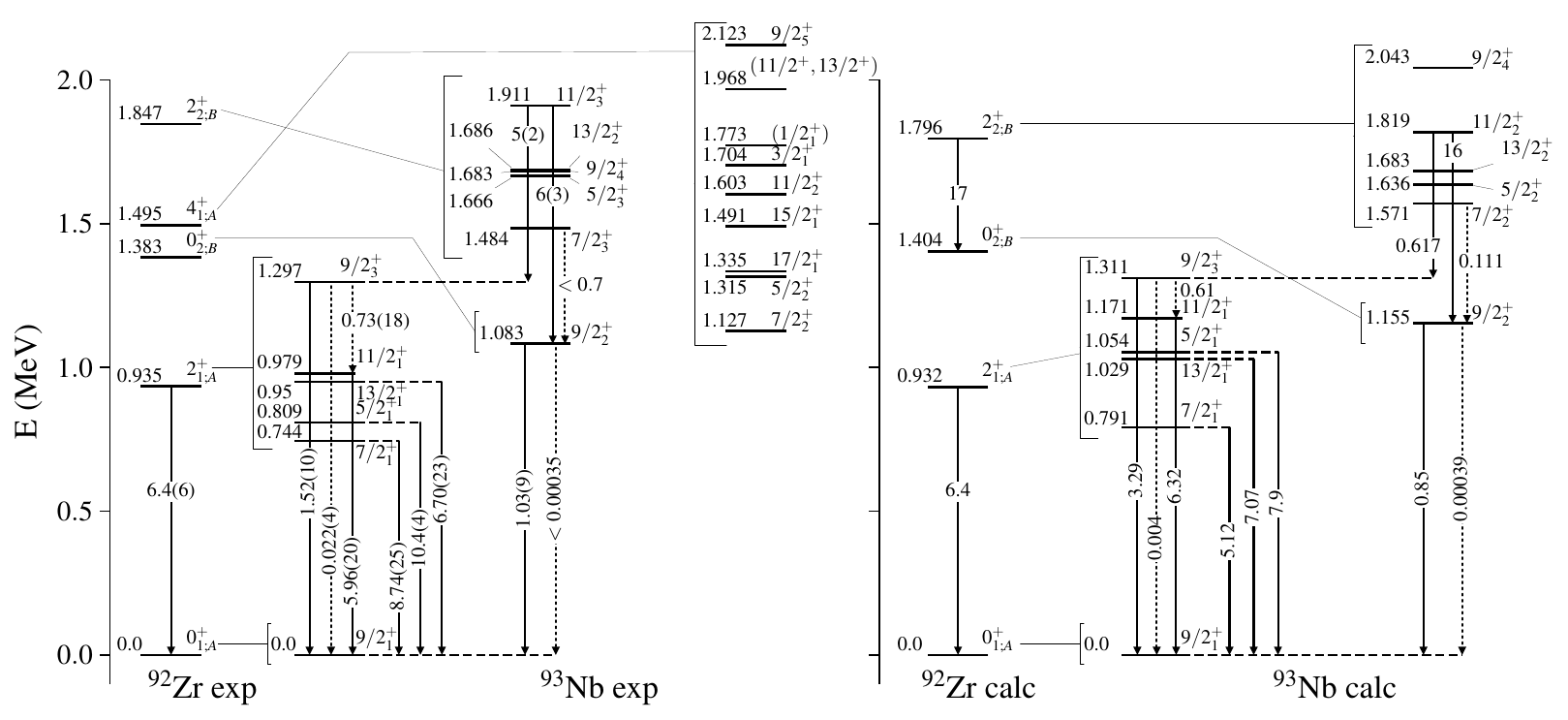}
\caption{Experimental (left) and calculated (right) energy 
levels in MeV, and $E2$ (solid arrows) and $M1$ (dashed 
arrows) transition rates in W.u., for $^{93}$Nb and 
$^{92}$Zr. Lines connect $L$-levels in $^{92}$Zr to sets of 
$J$-levels in $^{93}$Nb, indicating the weak coupling 
$(L\otimes \tfrac{9}{2})J$.
For the data, see Fig.~3 of~\cite{gavleviac22}.
Note that the observed $4^+_{\rm 1;A}$ 
state in $^{92}$Zr is outside the $N\!=\!1$ model 
space.\label{fig:93Nb-p}}
\end{figure*}

Additional evidence for a \mbox{Type~I} QPT,
involving shape changes  
within the intruder B~configuration,
is obtained by examining the
individual structure of Nb isotopes at the end-points
of  the region considered. Figure~\ref{fig:93Nb-p}
displays the experimental and calculated levels in
$^{93}$Nb along with 
$E2$ and $M1$ transitions among them. The corresponding 
spectra of $^{92}$Zr, the even-even core, are also shown  
with an assignment of each level $L$ to the normal A or 
intruder B configurations, based on the analysis 
in~\cite{Gavrielov2022}, which also showed that the two 
configurations in $^{92}$Zr are spherical or 
weakly-deformed. It has long been known
that low-lying states of the A~configuration in $^{93}$Nb, 
can be interpreted in a weak coupling scheme, where the 
single-proton $\pi(1g_{9/2})$ state is coupled to 
spherical-vibrator states of the core. Specifically, for 
the $0^+_{1;A}$ ground state of $^{92}$Zr, this coupling 
yields the ground state $9/2^+_1$ of $^{93}$Nb. For 
$2^+_{1;A}$, it yields a quintuplet of states, $5/2^+_1, 
7/2^+_1, 9/2^+_3, 11/2^+_1,13/2^+_1$, whose ``center of  
gravity'' (CoG), is 0.976~MeV, in 
agreement with the observed energy 0.935~MeV of $2^+_1$
in $^{92}$Zr. The $E2$ transitions from the quintuplet  
states to the ground state are comparable in magnitude
to the $2^+_{1;A}\!\to\!0^+_{1;A}$ transition in $^{92}$Zr, 
except for $9/2^+_3$, whose decay is weaker.
A nonet of states built on  
$4^+_{1;A}$ can also be identified in the empirical 
spectrum of  $^{93}$Nb, with CoG of 1.591~MeV, close to 
1.495~MeV of $4^+_{1;A}$.
\begin{figure}[b]
\centering
\includegraphics[width=\linewidth]{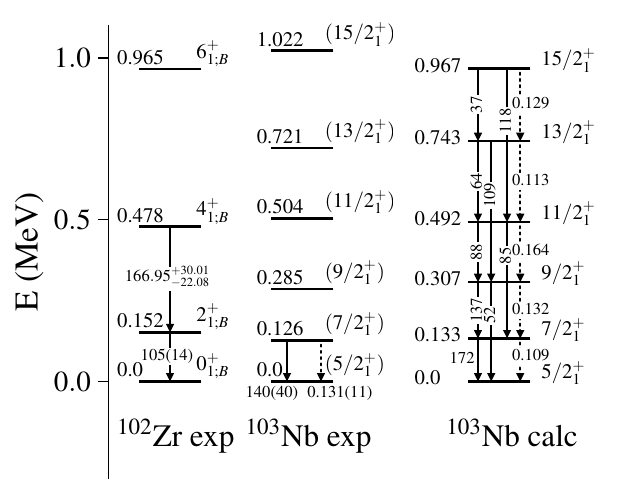}
\caption{Experimental and calculated energy 
levels in MeV, and $E2$ (solid arrows) and $M1$ (dashed 
arrows) transition rates in W.u., for $^{103}$Nb and
$^{102}$Zr. For the data,
see Fig.~4 of~\cite{gavleviac22}.
\label{fig:103Nb-p}}
\end{figure}

Particularly relevant to the present discussion is the
fact that the weak-coupling scenario is also valid for
non-yrast states of the intruder B configuration in
$^{93}$Nb. As shown in Fig.~\ref{fig:93Nb-p},
the coupling of $\pi(1g_{9/2})$ to 
the $0^+_{2;B}$ state in $^{92}$Zr, yields the excited 
$9/2^+_2$ state in $^{93}$Nb. For $2^+_{2;B}$~it yields the 
quintuplet, $5/2^+_3, 7/2^+_3, 9/2^+_4, 11/2^+_3, 
13/2^+_2$, whose CoG is 1.705~MeV, a bit lower than 
1.847~MeV of $2^+_{2;B}$. The observed $E2$ rates 
$1.03(9)$~W.u for \mbox{$9/2^+_2\!\to\!9/2^+_1$}, is
close to the calculated value 0.85~W.u., but is smaller
than the observed value $1.52(10)$~W.u for 
\mbox{$9/2^+_3\!\to\!9/2^+_1$}, suggesting that $9/2^+_2$ is 
associated with the B~configuration.

For $^{103}$Nb, the yrast states shown in
Fig.~\ref{fig:103Nb-p}
are arranged in a $K=5/2^+$ rotational band, with an 
established Nilsson model assignment 
$5/2^+[422]$. The band members can be interpreted in the 
strong coupling scheme, where a particle is coupled to an 
axially-deformed core. The indicated states are obtained by 
coupling the $\pi(1g_{9/2})$ state to the ground band 
($L=0^+_{1;B},2^+_{1;B},4^+_{1;B},
6^+_{1;B},\ldots$) of $^{102}$Zr, also shown in
Fig.~\ref{fig:103Nb-p},
which is  
associated with the intruder B~configuration. The 
calculations reproduce well the observed particle-rotor 
$J(J+1)$ splitting, as well as, the $E2$ and $M1$ 
transitions within the band. Altogether, we see an 
evolution of structure from weak-coupling of a spherical 
shape in $^{93}$Nb, to strong-coupling of a deformed shape 
in $^{103}$Nb. Such shape-changes within the 
B~configuration (Type~I QPT), superimposed on abrupt 
configuration crossing (Type-II QPT), are the key defining 
feature of intertwined QPTs. Interestingly, this  
intricate scenario, originally observed in the 
even-even Zr
isotopes~\cite{Gavrielov2019,Gavrielov2020,Gavrielov2022},
persists in the adjacent odd-even Nb~isotopes.

\section{Classical analysis}
One of the main advantages of the algebraic approach is
that one can do both a quantum and a classical analysis,
thus enabling an intuitive geometric interpretation. 
Geometry is introduced into the IBFM by means of
a matrix potential of the
Hamiltonian~(\ref{Hibfm})
in the basis~\cite{Lev88,LevShao89,Alonso92},
\ba
\ket{j,m;\beta,\gamma;N} =
\ket{j,m}\otimes\ket{\beta,\gamma;N} ~,
\ea
which involves the product of a $j$-fermion
$\ket{j,m}= a^{\dag}_{jm}\ket{0}$
and a projective coherent state
of $N$ bosons~\cite{GinoKir80,Diep80},
\ba
\ket{\beta,\gamma;N} = (N!)^{-1/2}(b^\dag_c)^N\ket{0} ~.
\ea
Here
${\small b^\dag_c \!=\!
(1+\beta^2)^{-1/2}
[\beta\cos\gamma d^\dag_0 +
  \tfrac{1}{\sqrt{2}}\beta\sin\gamma
  (d^\dag_2 + d^\dag_{-2}) + s^\dag]}$ and
$(\beta,\gamma)$ are quadrupole shape variables.
The resulting matrix, $\Omega_{N}(\beta,\gamma;\xi)$,
is real and symmetric with entries 
$E_{\rm b}(\beta,\gamma;N)\delta_{m_1,m_2} +
\epsilon_j\,\delta_{m_1,m_2} + Ng_{m_1,m_2}(\beta,\gamma)$.
Here
$E_{\rm b}(\beta,\gamma;N)$ is the expectation value
of $\hat{H}_{\rm b}$~(\ref{Hb})
in $\ket{\beta,\gamma;N}$,
\begin{eqnarray}
  \label{Eb-surface}
 &&
\hspace{-1cm}
E_{\rm b}(\beta,\gamma;\epsilon_d,\kappa,\chi;N) =
    5\kappa\, N +
    \tfrac{N\beta^2}{1+\beta^2} 
\left[\epsilon_d + \kappa (\chi^2-4)\right]
\nonumber\\
&&
\quad
\hspace{-1cm}
+ \tfrac{N(N-1)\beta^2}{(1+\beta^2)^2}\kappa
\left[4 - 4\bar{\chi}\beta\,\cos3\gamma
  + \bar\chi^2\beta^2\right] ,\qquad
\end{eqnarray} 
where $\bar\chi\!=\!\sqrt{\frac{2}{7}}\chi$.
Explicit expressions of
$g_{m_1,m_2}(\beta,\gamma)$
are given in Eqs.~(10)-(14) of~\cite{Petrellis2011b}.
Diagonalization of
the matrix splits into
two (doubly degenerate) pieces with
$m=j,j-2,\ldots, -(j-1)$ and similarly with $m\to -m$.
The dimension of the basis is thus $(j+1/2)$.
The resulting eigenvalues are the
Bose-Fermi potential surfaces
$E^{(N)}_k(\beta,\gamma;\xi)$,
which include the contribution of the core and of the
single particle levels in the deformed $\beta$ and
$\gamma$ field generated by the bosons.

For $\gamma=0$ (axial shape),
the potential matrix $\Omega_{N}(\beta;\xi)$
is diagonal in the basis $\ket{j,K;\beta;N}$
with entries~\cite{Lev88},
\bsub
\ba
&&
\hspace{-1cm}
E^{(N)}_{K}(\beta;\xi) =
E_{\rm b}(\beta;N) +
\epsilon_j + N\lambda_{K}(\beta) ~,\\
&&
\hspace{-1cm}
         \lambda_{K}(\beta;\chi,A,\Gamma,\Lambda) =
A\,\tfrac{\beta^2}{1+\beta^2}
\nonumber\\
&&
\quad
\hspace{-1cm}
+ \tfrac{\beta}{1+\beta^2}C_{jK}\,
[\, \Gamma\sqrt{5}(\beta\bar{\chi}-2) 
  -\beta\Lambda(2j+1)C_{jK}\,] ~,\qquad
\label{tl-lamb}
\ea
\label{EK-N}
\esub
where
$C_{jK} = \tfrac{3K^2-j(j+1)}
{\sqrt{(2j-1)(2j+1)(2j+3)j(j+1)}}$.
The dependence of $\lambda_{K}(\beta)$ on $K^2$
reflects the double degeneracy ($K\to -K$)
mentioned above.
\begin{figure*}[t]
  \begin{center}
\begin{overpic}[width=0.31\linewidth, clip]{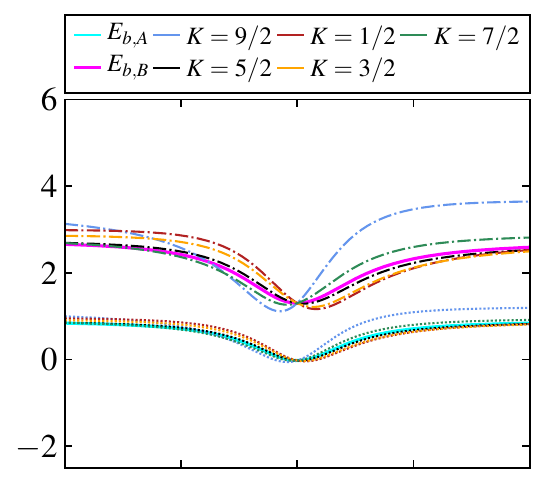}
\put(80,7) {$^{93}$Nb}
\put(15,7) {(a) $E_{A,B;K}(\beta)$}
\end{overpic}
\begin{overpic}[width=0.31\linewidth, clip]{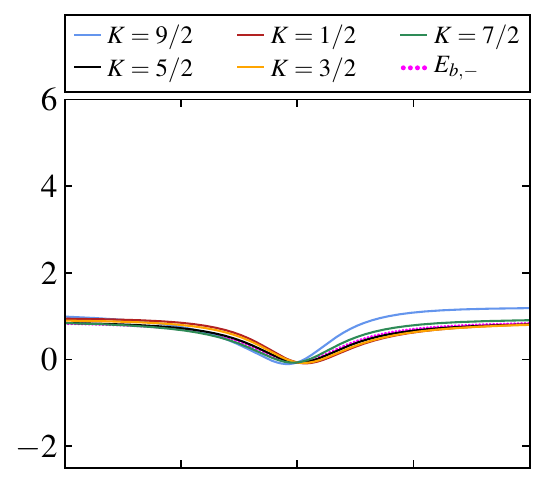}
\put(80,7) {$^{93}$Nb}
\put(15,7) {(b) $E_{-,K}(\beta)$}
\end{overpic}
\begin{overpic}[width=0.31\linewidth, clip]{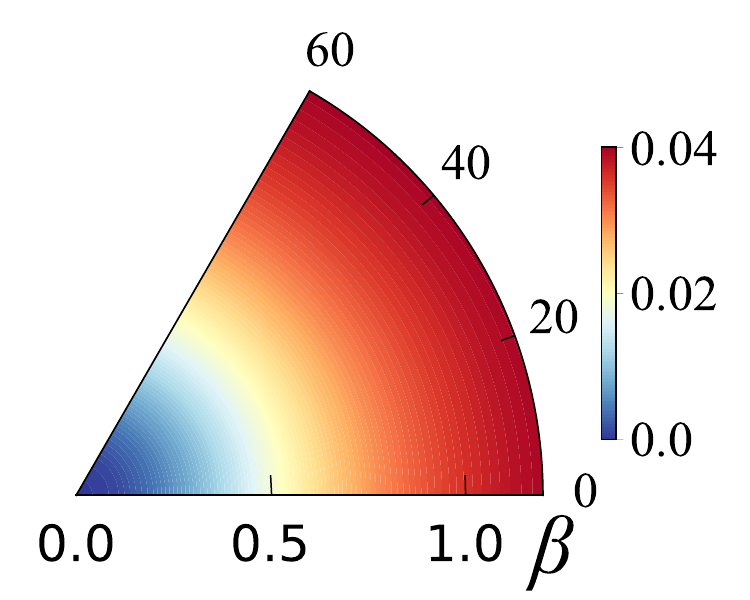}
\put(10,55) {\Large$^{92}$Zr}
\end{overpic}\\
\begin{overpic}[width=0.31\linewidth, clip]{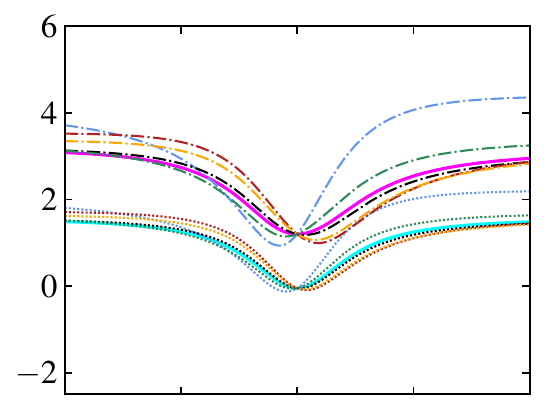}
\put(80,7) {$^{95}$Nb}
\put(15,7) {(c) $E_{A,B;K}(\beta)$}
\end{overpic}
\begin{overpic}[width=0.31\linewidth, clip]{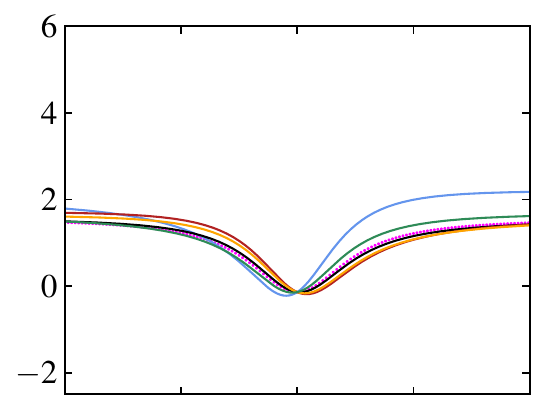}
\put(80,7) {$^{95}$Nb}
\put(15,7) {(d) $E_{-,K}(\beta)$}
\end{overpic}
\begin{overpic}[width=0.31\linewidth, clip]{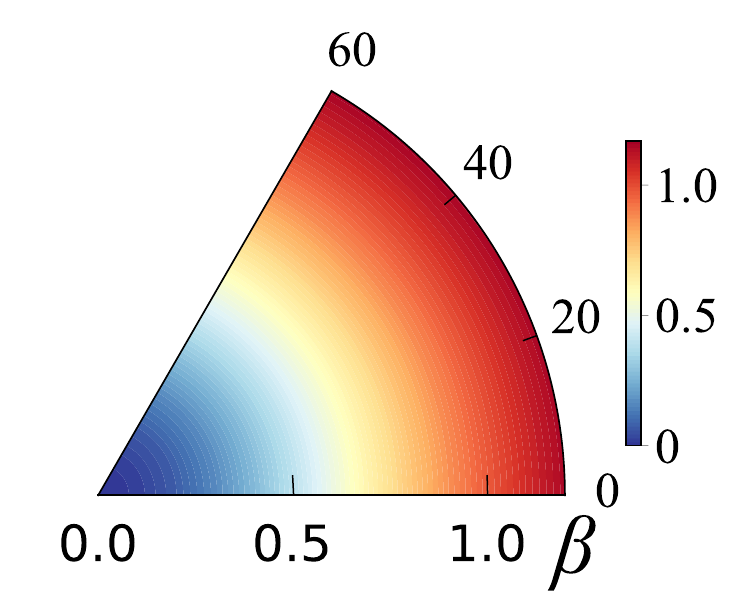}
\put(10,55) {\Large$^{94}$Zr}
\end{overpic}\\
\begin{overpic}[width=0.31\linewidth, clip]{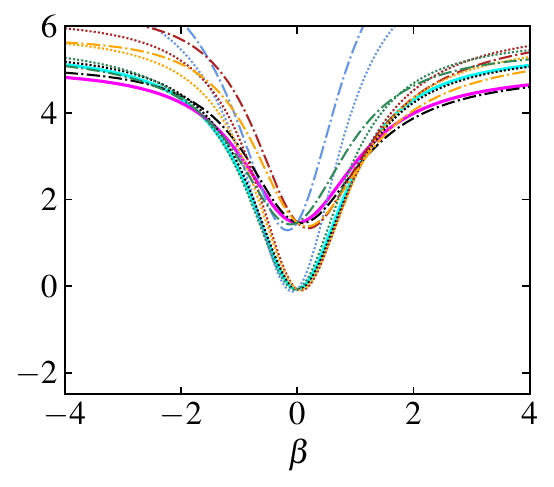}
\put(80,20) {$^{97}$Nb}
\put(15,20) {(e) $E_{A,B;K}(\beta)$}
\end{overpic}
\begin{overpic}[width=0.31\linewidth, clip]{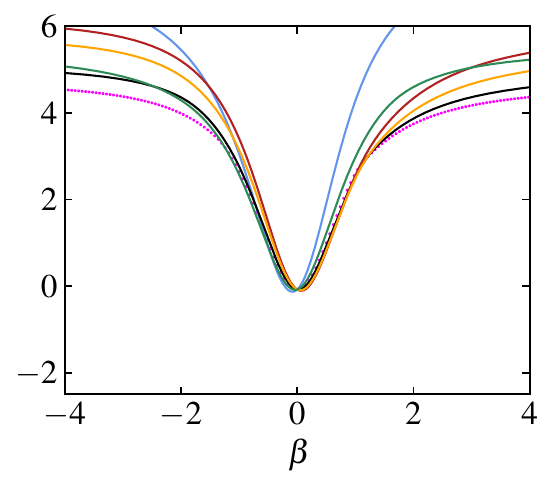}
\put(80,20) {$^{97}$Nb}
\put(15,20) {(f) $E_{-,K}(\beta)$}
\end{overpic}
\begin{overpic}[width=0.31\linewidth, clip]{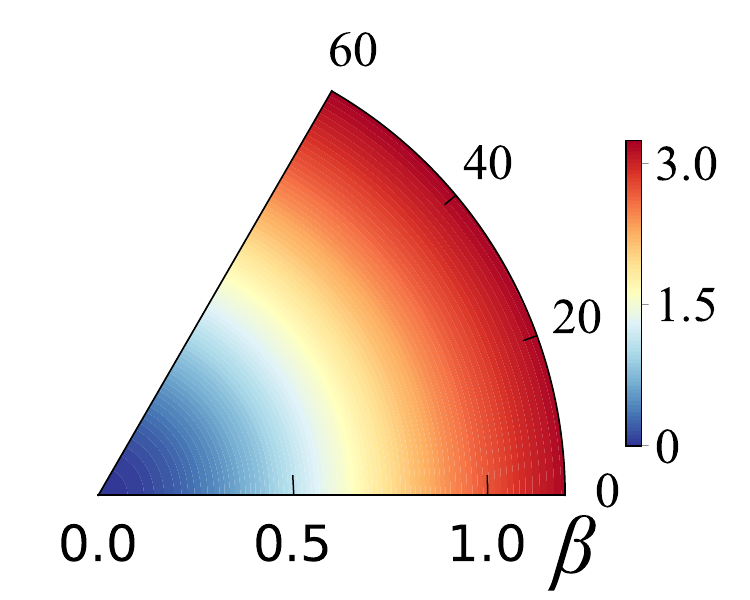}
\put(10,55) {\Large$^{96}$Zr}
\end{overpic}
\caption{\small
  Unmixed surfaces
  $E_{A,B;K}\equiv\{E_{A;K}(\beta),E_{B;K}(\beta)\}$,
  Eqs.~(\ref{EAK})-(\ref{EBK}),
  (left panels) and
  eigen-potentials $E_{-,K}(\beta)$,
  Eq.~(\ref{EpmK}), (middle panels)
  in MeV for $^{93,95,97}$Nb.
  Purely bosonic surfaces are also shown.
  Contour plots of eigen-potentials,
  $E_{-,K}(\beta,\gamma)$, for $^{92,94,96}$Zr (right~panels).
\label{Fig7}
}
\end{center}
\end{figure*}
\begin{figure*}[]
\begin{center}
\begin{overpic}[width=0.31\linewidth, clip]{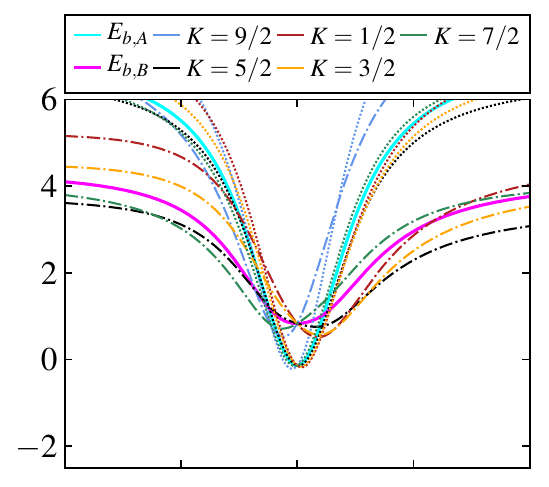}
\put(80,7) {$^{99}$Nb}
\put(15,7) {(a) $E_{A,B;K}(\beta)$}
\end{overpic}
\begin{overpic}[width=0.31\linewidth, clip]{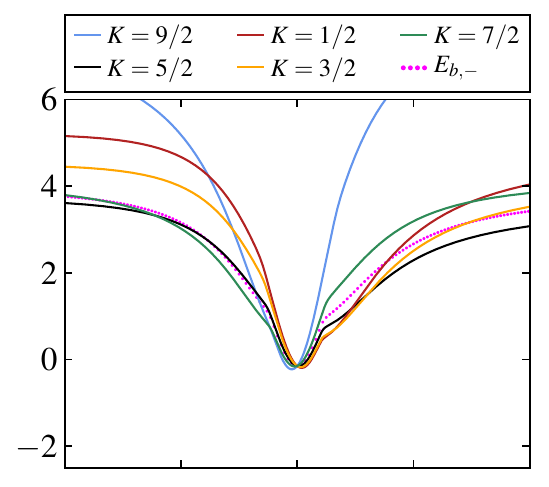}
\put(80,7) {$^{99}$Nb}
\put(15,7) {(b) $E_{-,K}(\beta)$}
\end{overpic}
    \begin{overpic}[width=0.31\linewidth, clip]{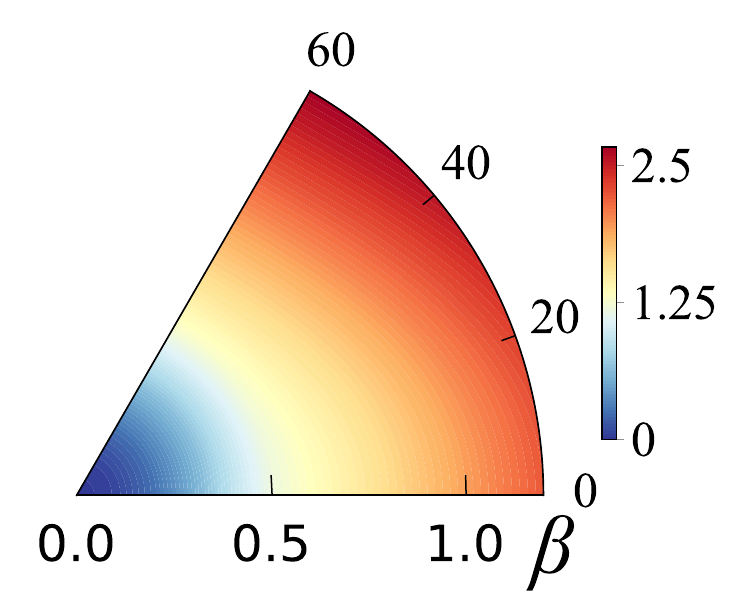}
\put(10,55) {\Large$^{98}$Zr}
\end{overpic}\\
    \begin{overpic}[width=0.31\linewidth, clip]{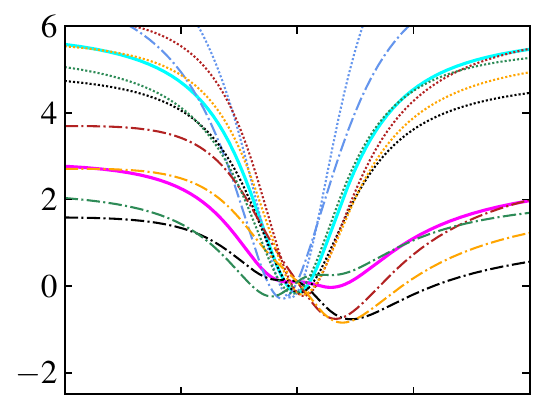}
\put(80,7) {$^{101}$Nb}
\put(15,7) {(c) $E_{A,B;K}(\beta)$}
\end{overpic}
    \begin{overpic}[width=0.31\linewidth, clip]{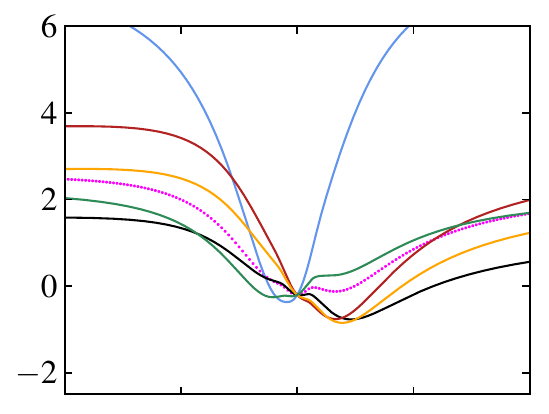}
\put(80,7) {$^{101}$Nb}
\put(15,7) {(d) $E_{-,K}(\beta)$}
\end{overpic}
    \begin{overpic}[width=0.31\linewidth, clip]{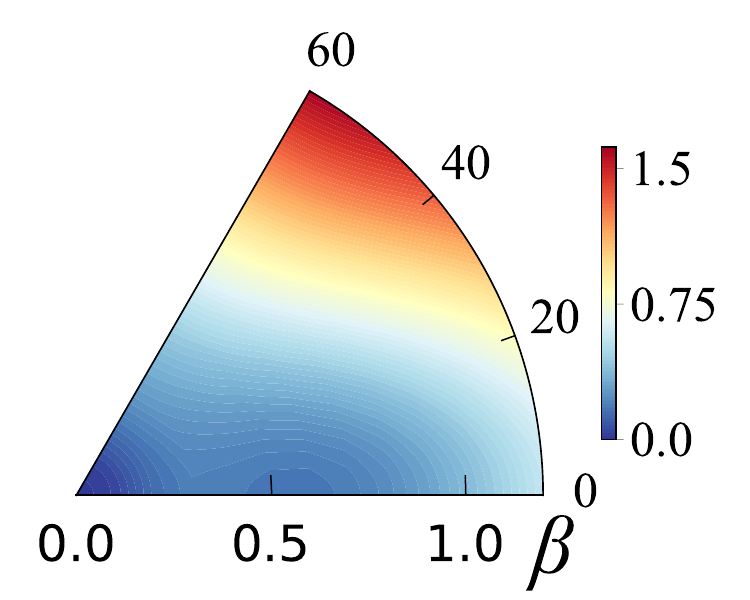}
\put(10,55) {\Large$^{100}$Zr}
\end{overpic}\\
    \begin{overpic}[width=0.31\linewidth, clip]{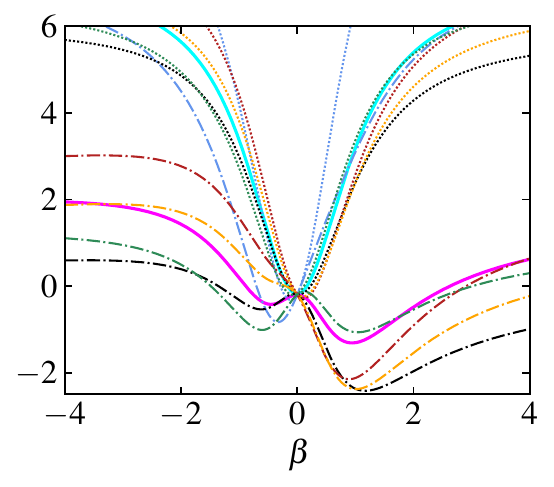}
\put(75,55) {$^{103}$Nb}
\put(15,20) {(e) $E_{A,B;K}(\beta)$}
\end{overpic}
    \begin{overpic}[width=0.31\linewidth, clip]{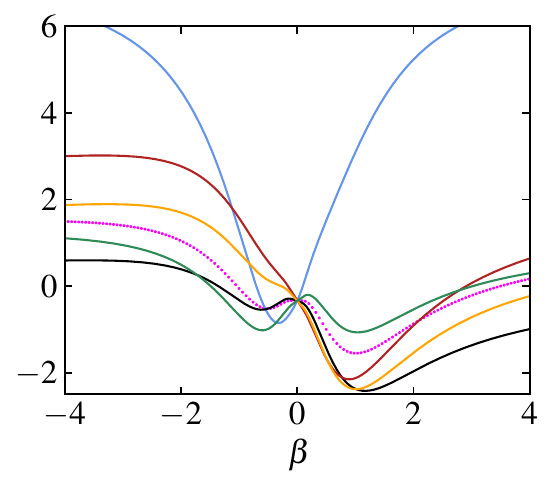}
\put(75,55) {$^{103}$Nb}
\put(15,20) {(f) $E_{-,K}(\beta)$}
\end{overpic}
    \begin{overpic}[width=0.31\linewidth, clip]{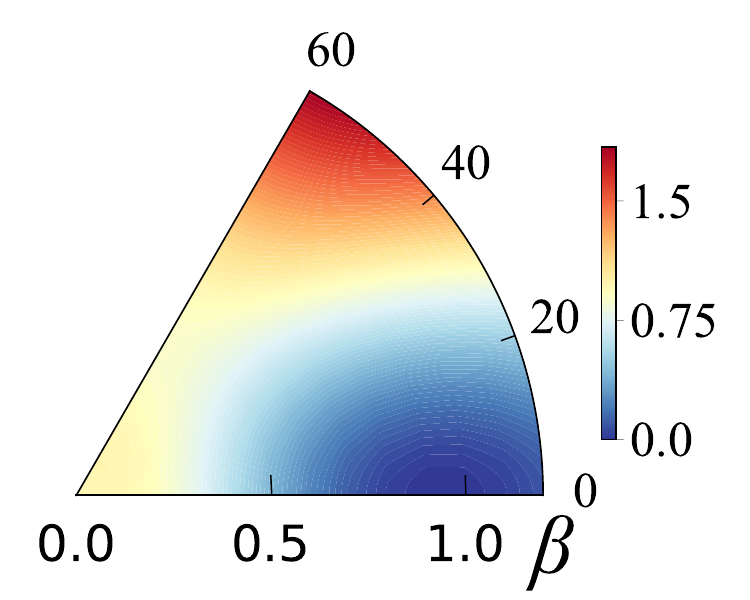}
\put(10,55) {\Large$^{102}$Zr}
\end{overpic}
\caption{\small
  Unmixed surfaces
  $E_{A,B;K}\equiv\{E_{A;K}(\beta),E_{B;K}(\beta)\}$,
  Eqs.~(\ref{EAK})-(\ref{EBK}),
  (left panels) and
  eigen-potentials $E_{-,K}(\beta)$,
  Eq.~(\ref{EpmK}), (middle panels)
  in MeV for $^{99,101,103}$Nb.
  Purely bosonic surfaces are also shown.
  Contour plots of eigen-potentials,
  $E_{-,K}(\beta,\gamma)$, for $^{98,100,102}$Zr
(right panels).
\label{Fig8}
}
\end{center}
\end{figure*}
\begin{figure*}[h]
\begin{center}
  \begin{overpic}[width=0.24\linewidth]{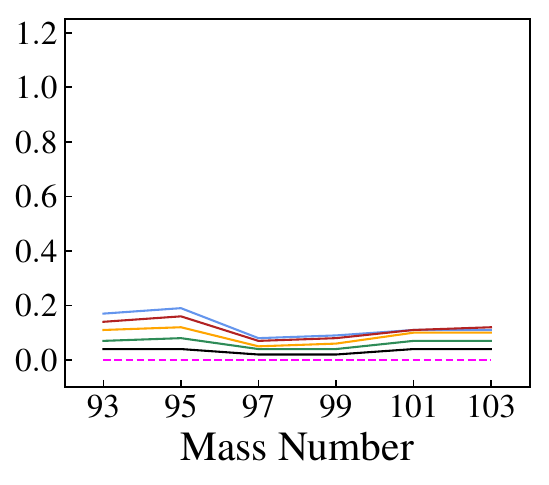}
\put(16,75) {(a) $\beta_{\rm eq}(E_{A;K})$}
\end{overpic}
  \begin{overpic}[width=0.24\linewidth]{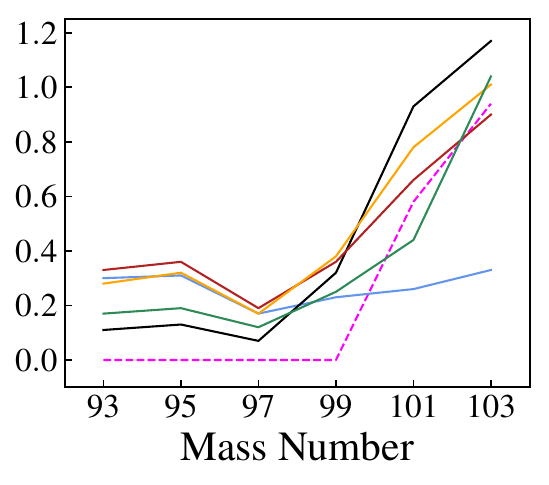}
\put(16,75) {(b) $\beta_{\rm eq}(E_{B;K})$}
\end{overpic}
  \begin{overpic}[width=0.24\linewidth]{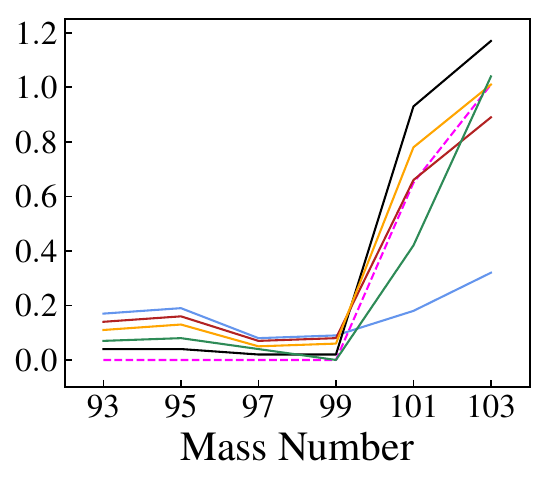}
\put(16,75) {(c) $\beta_{\rm eq}(E_{-,K})$}
\end{overpic}
  \begin{overpic}[width=0.24\linewidth]{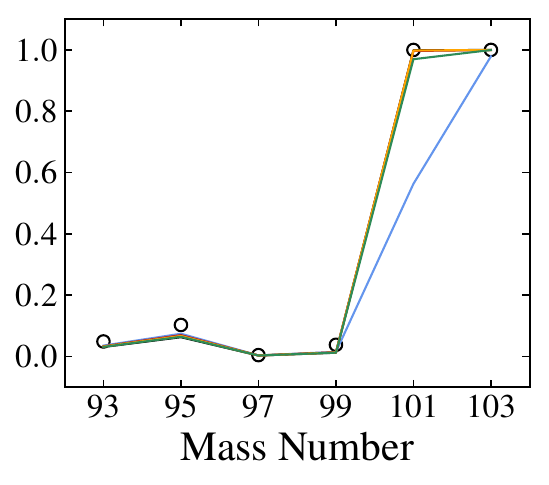}
\put(16,75) {(d) $b^2(K)$}
\end{overpic}
\caption{\small
  Equilibrium deformations (absolute value) of the
  unmixed surfaces
  (a)~$\beta_{\rm eq}(E_{A;K})$ and
  (b)~$\beta_{\rm eq}(E_{B;K})$,
  and of the lowest eigen-potential
  (c)~$\beta_{\rm eq}(E_{-,K})$.
  Dashed lines denote the minima,
  $\beta_{\rm eq}(E_{b,A}),\,\beta_{\rm eq}(E_{b,B}),
  \,\beta_{\rm eq}(E_{b,-})$, of the corresponding
  purely bosonic surfaces.
  (d)~Probability $b^2(K)$, Eq.~(\ref{b-prob}),
  of the intruder
  component in $\ket{\Psi_{-,K}(\beta)}$
  at $\beta_{\rm eq}(E_{-,K})$.
  Open circles ($\circ$)
  denote $b^2(J^{+}_{\rm gs})$ for the ground state in
  the quantum analysis~\cite{gavleviac22,Gav23},
  where $J^{+}_{\rm gs}\!=\!9/2^{+}$ ($5/2^{+}$)
for $^{93,95,97,99}$Nb ($^{101,103}$Nb).
Color coding for different
$K$ values as in Figs.~\ref{Fig7}-\ref{Fig8}.
\label{Fig9}
}
\end{center}
\end{figure*}

A geometric interpretation for the
IBFM-CM~\cite{LevGav25,Ramos25},
is obtained
by constructing from the Hamiltonian~(\ref{Hibfm-cm})
an enlarged potential matrix
of order $(2j+1)\times(2j+1)$, in the basis
$\{\ket{j,m_1;\beta,\gamma;N},
\ket{j,m_2;\beta,\gamma;N+2}\}$,
with $m_1,m_2 = j,j~-~2,\ldots, -(j-1)$,
\ba
\hspace{-0.5cm}
\Omega_{N}(y;\xi_A,\xi_B,\omega) =
\left [\begin{array}{c|c}
    \Omega_A(y;\xi_A) &
    \Omega_{AB}(y;\omega)\\
\hline
\Omega_{AB}(y;\omega) &
\Omega_B(y;\xi_B)
\end{array}
\right ] ~.
\label{Omega-cm}
\ea
The matrix depends on $y\equiv(\beta,\gamma)$ and on
the parameters of the Hamiltonian, Eq.~(\ref{Hibfm-cm2}), 
$\xi_A=(\epsilon^{A}_d,\kappa_{A},\chi,
\epsilon^{A}_j,A,\Gamma,\Lambda)$,
$\xi_B=(\epsilon^{B}_d,\kappa_{B},\chi,
\kappa^{\prime}_{B},\Delta_B,
\epsilon^{B}_j,A,\Gamma,\Lambda)$, $\omega$ and $N$.
The sub-matrix $\Omega_A(\beta,\gamma;\xi_A)$
acts in the
$\ket{j,m_1;y;N}$ sector, 
$\Omega_B(\beta,\gamma;\xi_B)$ acts in the 
$\ket{j,m_2;y;N+2}$ sector and
$\Omega_{AB}(\beta,\gamma;\omega)$ connects the two
sectors. For $\hat{H}_{\rm f}=\hat{V}_{\rm bf} =0$,  
the above potential matrix reduces to that proposed
for the IBM-CM~\cite{Vargas04}.
Diagonalization of the matrix
$\Omega_N(y;\xi_A,\xi_B,\omega)$ produces
the eigen-potentials.

For $\gamma\!=\! 0$,
the matrix~$\Omega_N(\beta;\xi_A,\xi_B,\omega)$
of Eq.~(\ref{Omega-cm}), can be transformed into
a simple block-diagonal form
$\{M_{K=j}(\beta),\,M_{K=j-2}(\beta),\,\ldots,\,
M_{K=-(j-1)}(\beta)\}$, where
$M_K(\beta)$ stands for a $2\times 2$ matrix in the
states $\ket{\Psi_{A;K}}\equiv\ket{j,K;\beta;N}$ and 
$\ket{\Psi_{B;K}}\equiv\ket{j,K;\beta;N+2}$,
\ba
M_K(\beta) =
\left [
\begin{array}{cc}
E_{A;K}(\beta) & W(\beta) \\
W(\beta) & E_{B;K}(\beta)
\end{array}
\right ] ~.
\label{MK}
\ea
Using Eq.~(\ref{EK-N}), the entries of $M_K(\beta)$ are
given by,
\bsub
\ba
&&
\hspace{-1cm}
E_{A;K}(\beta) =
E^{(N)}_{K}(\beta;\xi_A) ~,
\label{EAK}\\
&&
\hspace{-1cm}
E_{B;K}(\beta) =
E^{(N+2)}_{K}(\beta;\xi_B)
+ 6\kappa^{\prime}_{B}\,\tfrac{(N+2)\beta^2}{1+\beta^2}
+ \Delta_B ~,
\label{EBK}\\
&&
\hspace{-1cm}
W(\beta) =
\tfrac{\sqrt{(N+2)(N+1)}}{1+\beta^2}\omega\,
( 1 + \tfrac{1}{\sqrt{5}}\beta^2) ~.
\label{Wbeta}
\ea
\esub

Introducing the quantities,
\bsub
\ba
\delta_{K}(\beta) &=&
E_{B;K}(\beta) - E_{A;K}(\beta) ~,
\label{delK}\\[1mm]
R_{K}(\beta)
&=& \tfrac{\delta_{K}(\beta)}{2W(\beta)} ~,
\label{RK}
\ea
\label{delK-RK}
\esub
the eigen-potentials then read,
\ba
&&
\hspace{-1cm}
E_{\pm,K}(\beta) =
\tfrac{1}{2}\Sigma_{K}(\beta)
\pm \big|W(\beta)\big|\sqrt{1+[R_{K}(\beta)]^2} ~,\quad
\label{EpmK}
\ea
where $\Sigma_{K}(\beta) = E_{A;K}(\beta) + E_{B;K}(\beta)$,
and the corresponding eigenvectors are
\bsub
\ba
\ket{\Psi_{-,K}(\beta)} &=& a\,\ket{\Psi_{A;K}}
+ b\, \ket{\Psi_{B;K}} ~,
\label{Psi-m}
\\
\ket{\Psi_{+,K}(\beta)} &=& -b\,\ket{\Psi_{A;K}}
+ a\, \ket{\Psi_{B;K}} ~. 
\label{Psi-p}
\ea
\label{Psi-pm}
\esub
The ratio of mixing amplitudes and probabilities satisfy,
\bsub
\ba
&&\frac{b}{a} = \left [ R_{K}(\beta)
  \mp \sqrt{1+[R_{K}(\beta)]^2}\right ] ~,\\
&&b^2 = 1 - a^2 =\tfrac{1}{1+ \Bigl [ \,R_{K}(\beta)\,
    \pm \sqrt{1+[R_{K}(\beta)]^2}\,\Bigr ]^2} ~,
\label{b-prob}
\ea
\label{a-b-prob}
\esub
where the upper (lower)
sign applies for $W(\beta)>0$
[$W(\beta)<0]$.

The ensemble of
eigen-potentials~$\{E_{-,K}(\beta),\,E_{+,K}(\beta)\}$,
Eq.~(\ref{EpmK}), portray the change in energies of the odd
fermion as a function of deformation $\beta$, 
in the presence of coupled bosonic cores.
The states
$\{\Psi_{-,K}(\beta),\,\Psi_{+,K}(\beta)\}$,
Eq.~(\ref{Psi-pm}), depict the change in
configuration content.
The value of $R_{K}(\beta)$, Eq.~(\ref{RK}),
determines the character of the
normal-intruder mixing. We observe maximal mixing for
$R_K(\beta)\!=\!0$, strong mixing for
$|R_K(\beta)|\!<<\!1$,
weak mixing for $|R_K(\beta)|\!>>\!1$ and no mixing for
$|R_K(\beta)|\!=\!\infty$.

Two scenarios are relevant for the subsequent discussion.
(a)~Maximal mixing occurs for $\delta_{K}(\beta)\!=\!0$,
{\it i.e.}, when the two unmixed surfaces are degenerate,
$E_{A;K}(\beta)=E_{B;K}(\beta)\equiv E^{(0)}_K(\beta)$.
In this case, the eigen-potentials are
$E_{\pm}(\beta)= E^{(0)}_K(\beta) \pm |W(\beta)|$ and the
corresponding eigenfunctions, Eq.~(\ref{Psi-pm}),
have $\tfrac{b}{a} = +1$ ($-1$) for
$W(\beta)<0$ [$W(\beta)>0$]. 
(b)~No mixing occurs for $W(\beta)=0$.
  In this case, for $\delta_K(\beta)>0$:
  $E_{-,K}(\beta) = E_{A;K}(\beta)$,
  $\ket{\Psi_{-,K}}=\ket{\Psi_{A;K}}$, 
  $E_{+,K}(\beta) = E_{B;K}(\beta)$,
  $\ket{\Psi_{+,K}}=\ket{\Psi_{B;K}}$.
  In contrast, for $\delta_K(\beta)<0$:
  $E_{-,K}(\beta) = E_{B;K}(\beta)$,
 $\ket{\Psi_{-,K}} = \ket{\Psi_{B;K}}$, 
 $E_{+,K}(\beta) = E_{A;K}(\beta)$,
 $\ket{\Psi_{+,K}} = -\ket{\Psi_{A,K}}$.
  In what follows, we apply the formalism to interpret
  geometrically the results of the quantum analysis
  in~Section~\ref{sec-Nb} for odd-mass
  Nb isotopes~\cite{LevGav25}.

Intertwined QPTs are characterized by a weak coupling
[small $W(\beta)$, Eq.~(\ref{Wbeta})]
and a rapid crossing of the two configurations.
At the crossing point
[$\delta_{K}(\beta)\!=\!0$, Eq.~(\ref{delK})],
the mixing in $\ket{\Psi_{\pm,K}}$, Eq.~(\ref{Psi-pm}),
is maximal.
The crossing of $E_{A;K}(\beta)$ and $E_{B;K}(\beta)$
implies a change in sign of $\delta_K(\beta)$,
and the two eigenfunctions
$\{\Psi_{-,K},\, \Psi_{+,K}\}$
interchange their character. Away from the crossing point,
the system rapidly converges to the no-mixing scenario
mentioned above.

Figures \ref{Fig7}-\ref{Fig8} show
for each $K$ the unmixed surfaces,
$E_{A,B;K}\equiv\{E_{A;K}(\beta),\,E_{B;K}(\beta)\}$
(left panels) and the lowest eigen-potentials,
$E_{-,K}(\beta)$ (middle panels)
which serve as the Landau potentials
for the Nb isotopes considered.
Also shown are the purely bosonic surfaces
$\{E_{\rm b,A}(\beta),\,E_{\rm b,B}(\beta)\}$,
and $E_{\rm b,-}(\beta)$, obtained by taking
$\hat{H}_{\rm f}\!=\!\hat{V}_{\rm bf}\!=\!0$ in the
Hamiltonian~(\ref{Hibfm-cm2}).
The contour plots $E_{\rm b,-}(\beta,\gamma)$
(right panels) are the lowest eigen-potentials
of the adjacent even-even Zr cores.
The latter classical potentials confirm the quantum
results~\cite{Gavrielov2019,Gavrielov2020,Gavrielov2022},
as they show a transition from spherical
($^{92-98}$Zr), to a flat-bottomed potential at $^{100}$Zr,
and to a prolate axially-deformed ($^{102}$Zr).
For $\beta\!=\!0$, the unmixed surfaces,
$E_{A,B;K}(\beta)$,
are independent of $K$. For $\beta\!\neq\! 0$, they
  exhibit quadratic and quartic $K$-splitting,
  Eq.~(\ref{tl-lamb}). The resulting landscape
  is asymmetric in $\beta$ and
is identical to that encountered in the IBFM with
a single configuration~\cite{Lev88}.
As discussed below, the presence or absence of
crossing of the unmixed surfaces, has
a direct impact on the
topology of the eigen-potentials
$E_{\pm,K}(\beta)$, Eq.~(\ref{EpmK}).

In $^{93,95}$Nb, we see from Figs.~\ref{Fig7}(a) and
\ref{Fig7}(c), that
$\delta_{K}(\beta)\!>\!0$ for all values of $\beta$,
hence for each $K$,
$E_{A;K}(\beta)$ and $E_{B;K}(\beta)$,
are well separated and do not intersect.
Consequently, the eigen-potentials, 
shown in Figs.~\ref{Fig7}(b) and \ref{Fig7}(d),
are similar to the
unmixed surfaces,
$E_{-,K}(\beta) \approx E_{A;K}(\beta),\,
E_{+,K}(\beta) \approx E_{B;K}(\beta)$.

In $^{93}$Nb,
all $K$-surfaces $E_{-,K}(\beta)$ of Fig.~\ref{Fig7}(b),
are close in energy and are similar to the boson surface,
$E_{\rm b,-}(\beta)$, with a minimum at $\beta\!=\!0$.
This behavior reflects a spherical core shape weakly
coupled to a $j$-fermion, consistent with
the quantum analysis of Section~\ref{sec-Nb}.
The latter assigns a weak-coupling type of
wave function $\ket{(L\otimes j)J}$
with $L\!=\!0,\,j\!=\!J\!=\!9/2$,
to the normal ground state~\cite{gavleviac22,Gav23}.

In $^{95}$Nb,
$E_{-,K}(\beta)$ of Fig.~\ref{Fig7}(d), display similar
topology but are more dispersed.
For small $\beta$, the
observed $K$-splitting is linear in $\beta$ and
quadratic in $K$,
in accord with Eq.~(\ref{tl-lamb}).
The factor $C_{jK}\propto [3K^2-j(j+1)]$ implies
opposite shifts for $K=1/2,\,3/2,\,5/2$ and
$K=7/2,\,9/2$ levels, with respect
to the boson surface $E_{\rm b,-}(\beta)$.

In $^{97,99,101,103}$Nb, we see from
Figs.~\ref{Fig7}(e), \ref{Fig8}(a), \ref{Fig8}(c),
\ref{Fig8}(e),
the occurrence of regions in $\beta$
with different signs for $\delta_{K}(\beta)$,
Eq.~(\ref{delK}),
due to crossing of the unmixed surfaces.
The crossing points
are on the prolate side at $\beta^{*}_K>0$ and
on the oblate side at $\beta^{**}_K<~0$. At these points,  
$\delta_{K}(\beta^{*}_K)=\delta_{K}(\beta^{**}_K)=0$
and $\ket{\Psi_{\pm,K}}$, Eq.~(\ref{Psi-pm}),
exhibit maximal mixing.
$\beta^{**}_K$ and $\beta^{*}_K$
mark the borders of regions with
alternating sign of $\delta_{K}(\beta)$.
I)~$\beta < \beta^{**}_K$ ($\delta_{K}(\beta) < 0$);
II)~$\beta^{**}_K < \beta < \beta^{*}_K$ 
($\delta_{K}(\beta) > 0$);
III)~$\beta > \beta^{*}_K$
($\delta_{K}(\beta) < 0$).
Region~II includes $\beta=0$.
As the system evolves from one region to adjacent~one, 
the two eigen-potentials switch their
configuration-content ($A\!\leftrightarrow\! B$).
Specifically, in regions~I and III, we find
$E_{-,K}(\beta) \approx E_{B;K}(\beta)$ and
$E_{+,K}(\beta) \approx E_{A;K}(\beta)$,
while in region~II,
$E_{-,K}(\beta) \approx E_{A;K}(\beta)$ and
$E_{+,K}(\beta) \approx  E_{B;K}(\beta)$.

In $^{97}$Nb,
the crossing of 
$E_{A;K}(\beta)$ and $E_{B;K}(\beta)$,
shown in Fig.~\ref{Fig7}(e),
occurs at high energies and their slopes
at the crossing points ($\beta^{*}_K$ and $\beta^{**}_K$)
are similar.
Consequently, these crossings have little effect on the
eigen-potentials, shown in Fig.~\ref{Fig7}(f).
In particular, the minimum of
$E_{-,K}(\beta)$ remains at $\beta=0$,
and in its vicinity $E_{-,K}(\beta)\approx E_{A;K}(\beta)$.

In $^{99}$Nb, the crossing of 
unmixed surfaces, shown in Fig.~\ref{Fig8}(a),
occurs at lower energies, and their slopes
at the crossing points are different.
This leads to ``kinks'' in the eigen-potentials
of Fig.~\ref{Fig8}(b).
Particularly noticeable are the kinks in the
$K=1/2,\,3/2,\,5/2$ levels, exhibiting a downward bend
in $E_{-,K}(\beta)$ on the prolate side.
Such kinks signal the
approach to the critical point of the Type~II QPT at
neutron number~60,
where the ground state changes from the normal
($A$) to the
intruder ($B$)~configuration. The minimum of
$E_{-,K}(\beta)\approx E_{A;K}(\beta)$
is still at $\beta=0$.

The crossing points
$\beta^{*}_K \!>\!0$ and $\beta^{**}_K\!<\!0$ of the
unmixed surfaces
approach each other as the mass
number increases. Consequently, region~II
shrinks and in the heavier $^{101,103}$Nb isotopes,
the eigen-potentials satisfy
$E_{-,K}(\beta)\approx E_{B;K}(\beta)$
for most values of~$\beta$.

In $^{101}$Nb, the $E_{-,K}(\beta)$ surfaces with
$K\!=\!1/2,3/2,5/2$, develop
prolate-deformed minima in region~III,
reflecting a transition to rotational-band structure of
intruder $K$-bands. As seen in Fig.~\ref{Fig8}(d),
these deformed minima are deeper than the shallow minimum
in the flat-bottomed boson surface $E_{\rm b,-}(\beta)$
of $^{100}$Zr, and occur at different locations. 
This highlights the effect
of the odd fermion on the QPT in the vicinity
of the critical point.

In $^{103}$Nb, all surfaces satisfy
$E_{-,K}(\beta)\!\approx\!E_{B;K}(\beta)$
and support pronounced prolate-deformed minima
(except for $K\!=\!9/2$), upon which rotational $K$-bands
of intruder states are built.
The lowest bandhead in Fig.~\ref{Fig8}(f) has $K\!=\!5/2$,
in line
with the quantum analysis of Section~\ref{sec-Nb}.
The latter assigns a strong coupling
type of wave function to members of the ground
band~\cite{gavleviac22,Gav23}.
The surfaces 
$E_{-,K}(\beta)$ with $K\!=\!5/2,7/2$, support also
oblate-deformed local minima.

The equilibrium deformations obtained from the
global minimum of the unmixed surfaces
and lowest eigen-potential,
$\beta_{\rm eq}(E_{A;K}),\,\beta_{\rm eq}(E_{B;K})$ 
and $\beta_{\rm eq}(E_{-,K})$, serve as order
parameters of the QPT. Their evolution along the
Nb chain is shown in Fig.~\ref{Fig9}(a), \ref{Fig9}(b),
\ref{Fig9}(c),
along with
$\beta_{\rm eq}(E_{b,A}),\,\beta_{\rm eq}(E_{b,B}),
\,\beta_{\rm eq}(E_{b,-})$ of the corresponding
boson surfaces. (For $^{101}$Nb,
$E_{b,-}(\beta)$ exhibits close-in-energy
spherical and deformed minima).
The normal $A$~configuration remains spherical along
the Nb chain ($\beq(E_{A;K}) \!<\! 0.2$),
while the
intruder $B$~configuration changes gradually from
weakly deformed in $^{93}$Nb
($\beq(E_{B;K})\!\approx\! \textstyle{0.3\!-\!0.4}$)
to strongly deformed in $^{103}$Nb
($\beq(E_{B.K})\!\approx\!\textstyle{0.9\!-\!1.2}$
(except for $K\!=\!9/2$).
$\beq(E_{-,K})$ is similar to
$\beq(E_A)$ for $^{93,95,97,99}$Nb
and coincides with $\beq(E_{B;K})$ for $^{101,103}$Nb.

Fig.~\ref{Fig9}(d) shows the probability $b^2(K)$ of
the intruder component in the eigenvector
$\ket{\Psi_{-,K}}$, Eq.~(\ref{Psi-m})
at $\beq(E_{-,K})$, along the Nb chain.
The rapid change in structure from the normal
$A$~configuration in $^{93-99}$Nb (small $b^2$) to the
intruder $B$~configuration in $^{101,103}$Nb (large $b^2$)
is clearly evident. The values of $b^2(K)$
calculated from Eq.~(\ref{b-prob}),
agree with the exact values of $b^2(J^{+}_{\rm gs})$
in the ground state, obtained from Eq.~(\ref{Prob-ab-cm})
and shown in Fig.~\ref{fig-b2-be2-q-mu}(a).
The combined results of Fig.~\ref{Fig9} 
confirm the scenario of intertwined QPTs
in the odd-mass Nb isotopes.
\vspace{-0.21cm}

\section{Concluding remarks}
\label{sec-concl}
We have investigated the evolution of structure in
odd-mass Nb isotopes in the framework of the
IBFM-CM. A~quantum analysis discloses the effects
on spectral properties of an abrupt crossing of
normal and intruder states (Type~II QPT)
accompanied by a gradual evolution from spherical- to
deformed-core shapes and transition from weak to
strong coupling within the intruder configuration
(Type~I QPT). The pronounced presence of both types of 
QPTs demonstrates the occurrence of
 intertwined QPTs in these odd-mass nuclei.
A classical analysis, based on an extended
matrix coherent states formalism,
clarifies the effects of deformation and
configuration-mixing on the single particle motion, and
captures essential features of the quantum results.

\begin{acknowledgement}
A fruitful collaboration with N. Gavrielov (HU)
and F. Iachello (Yale) on the topics considered,
is acknowledged.
\end{acknowledgement}

\let\clearpage\relax

\end{document}